\documentclass[prl,twocolumn,amsmath,amssymb,superscriptaddress]{revtex4}
\usepackage{graphicx}% Include figure files

\begin{document}

\title{Charge Hall effect driven by spin-dependent chemical potential gradients and
 Onsager relations in mesoscopic systems}
\author{E. M. Hankiewicz}
\affiliation{Department of Physics, Texas A\&M University, College
Station, TX 77843-4242}
\author{Jian Li}
\affiliation{Department of Physics, The University of Hong-Kong,
Pukfulam Road, Hong-Kong, China}
\author{Tomas Jungwirth}
\affiliation{Department of Physics, University of Texas, Austin,
TX 78712-0264}
\author{Qian Niu}
\affiliation{Department of Physics, University of Texas, Austin,
TX 78712-0264}
\author{Shun-Qing Shen}
\affiliation{Department of Physics, The University of Hong-Kong,
Pukfulam Road, Hong-Kong, China}
\author{Jairo Sinova}
\affiliation{Department of Physics, Texas A\&M University, College
Station, TX 77843-4242}

\date{\today}

\begin{abstract}
We study theoretically the spin-Hall effect as well as its
reciprocal phenomenon (a transverse charge current driven by a
spin-dependent chemical potential gradient) in electron and hole
finite size mesoscopic systems. The Landauer-Buttiker-Keldysh
formalism is used to model samples with mobilities and Rashba
coupling strengths which are experimentally accessible and to
demonstrate the appearance of measurable charge currents induced
by the spin-dependent chemical potential gradient in the
reciprocal spin-Hall effect. We also demonstrate that within the
mesoscopic coherent transport regime the Onsager relations are
fulfilled for the disorder averaged conductances
for electron and hole mesoscopic systems.
\end{abstract}

\maketitle

{\it Introduction.} In the very active field of semiconductor
based spintronics the control of spin can be achieved by the
manipulation of the strength of spin-orbit (SO) interactions in
paramagnetic systems. Within this context, the newly proposed
intrinsic spin-Hall effect (SHE) in p-doped semiconductors  by
Murakami {\it et al}.\cite{Murakami03} and in a two-dimensional
electron  system (2DES) by  Sinova {\it et al.} \cite{sinova04}
offers new possibilities for  spin current manipulation and
generation in high  mobility paramagnetic semiconductor systems.
In contrast to the earlier proposed  extrinsic spin-Hall effect
\cite{Perel,Hirsch99,Zhang00}, which is associated with scattering
from impurities, the intrinsic spin-Hall arises purely from host
semiconductor band structure and represents a spin-current
response generated perpendicular to the driving electric field.

Recently, the spin Hall effect was experimentally observed by Kato
{\it et al.} \cite{Kato04} in n-doped GaAs using the Kerr effect
and by Wunderlich {\it et al.} \cite{Wunderlich} in the p-n
junction light-emitting diodes based on two dimensional hole gas
(2DHG)(Al,Ga)As. Although, the experiment by Wunderlich {\it et
al.} seems to be in the regime where the intrinsic effect in 2DHG
is dominant, the main theoretical focus has been concentrated so
far on 2DEG with Rashba SO interactions, where Rashba term is
linear with k
\cite{sinova04,All_part,Rashba03,Rashba04,Dimitrova04,Halperin04,
kentaro04,Inoue04,Chalaev04,Khaetskii04,Nikolic04,Sheng04,Hankiewicz04,
Murakamic05,Murakami05r}.The influence of disorder on infinite
2DEG is still unclear(for a  recent review see
\cite{Murakami05r}). For $\delta$-function impurities the
analytical calculations of vertex corrections in the ladder
approximation seem to cancel the intrinsic spin-Hall effect in a
weak scattering regime \cite{Inoue04,Halperin04}. However,these
calculations have been challenged recently \cite{Murakamic05}.
Furthermore, the numerical calculations  based on Kubo formula
using continuum model in momentum space \cite{kentaro04} and
discrete model in the real space \cite{sheng05} show finite
spin-conductivity in a weak scattering regime which goes to zero
in the thermodynamic limit for electron systems
\cite{Nomuraun,sheng05}. In contrast, for the infinite 2DHG the
vertex corrections vanish \cite{MurakamiPRB04,Bernevig04a}.
Moreover, the numerical calculations  based on Kubo formula using
continuum model in momentum space  show finite value of SHE in a
weak scattering regime which goes to constant in the thermodynamic
limit \cite{Nomuraun}.

 The calculations within the
Landauer-Buttiker (LB) formalism on finite size systems
\cite{Nikolic04,Sheng04,Hankiewicz04,Shen05} model a  sample of
micro/nanosize attached to contacts. The calculations on electron
mesoscopic systems show that spin Hall conductance is a fraction
of $e/8\pi$ in a weak scattering regime
\cite{Nikolic04,Sheng04,Hankiewicz04}. Moreover, a mesoscopic
spin-Hall conductance is robust against the disorder
\cite{Nikolic04,Sheng04,Hankiewicz04}. Very recently Wu and Zhou
considered the Luttinger model \cite{Wu05}, showing as expected
that SHE can be much larger in hole systems in comparison with the
electron ones. Although the experimental measurement by Wunderlich
at al \cite{Wunderlich} concern 2DHG systems with broken inversion
symmetry, the pure cubic Rashba term was not considered in detail
within the LB formalism so far.

The observation of spin-Hall effect through transport measurements
is one of the urgent experimental challenges facing this
spin-transport physics. Recently, an H-probe structure has been
proposed to measure the effect where the spin-Hall effect could be
measured indirectly by detecting charge voltages induced by the
reciprocal spin-Hall effect (RSHE) \cite{Hankiewicz04}. This RSHE,
where transverse charge current is driven by spin dependent
chemical potential, was proposed in a context of extrinsic
spin-Hall effect by Hirsch \cite{Hirsch99} and formulated in a
semiclassical approach by Zhang and Niu \cite{Niu04}. Also, the
Onsager relation between the spin-Hall conductivity and reciprocal
charge-Hall conductivity was established within a wave packet
model through a redefinition of the spin-current including
spin-torque terms in the bulk \cite{Niu04}. We show here that
within the mesoscopic regime, and more specifically within the
Landauer-Buttiker-Keldysh formalism, the Onsager relations are
satisfied within the models studied for the disorder averaged
conductances. Because the conductances are formulated with respect
to the leads which have no spin-orbit coupling, it is not
necessary nor consequential to introduce the spin current
redefinition in our problem \cite{Niu04}.

In this paper we  compare the magnitude of the SHE as well as the
RSHE in finite electron and hole mesoscopic systems within the LB
formalism.  We  show that the conductances associated with both
effects are significantly larger in the  hole systems.
Furthermore, we  analyze the possible experimental setup to
measure  the RSHE. We show that the charge current driven by a
spin-dependent chemical gradient is on the order of hundred
nano-ampers for  typical voltages in hole systems and should be
experimentally measurable.

{\it  Model  Hamiltonian for hole system and  LB   treatment  of
the spin-Hall effect and its reciprocal correspondent}.
 The observation of the spin-Hall effect and its reciprocal phenomenon
 in transport is the next experimental challenge in the subfield of spintronics
using spin-orbit interactions to manipulate the spin.

The continuum effective  mass  model for 2DHG in a narrow inversion asymmetrical well
 is given by
\cite{Schliemann05}: $\hat{H}=\frac{\hat{p}^2}{2m^*}
+i\frac{\lambda}{2\hbar^3}(p_{-}^{3}\hat{\sigma}_{+}-p_{+}^{3}\hat{\sigma}_{-})
+H_{dis}$, where $H_{dis}$ describes disorder. We use the
tight-binding approximation \cite{Datta} to model the disordered
conductor  within the  LB formalism. Within this approximation the
continuum  effective mass  envelope function
Hamiltonian becomes: %, Eq. \ref{Henv_func}, becomes:
\begin{eqnarray}
H&=&\sum_{j,\sigma} \epsilon_j c^{\dag}_{j,\sigma} c_{j,\sigma}
-t\sum_{j,\vec{\delta},\sigma}  c^{\dag}_{j+\vec{\delta},\sigma}
c_{j,\sigma}\nonumber \\ &+&t_{SO-k^3}[\sum_{j}
c^{\dag}_{j,\uparrow} c_{j+2a_{x},\downarrow}
-c^{\dag}_{j,\uparrow} c_{j-2a_{x},\downarrow}\nonumber \\
&+&i\sum_{j}c^{\dag}_{j,\uparrow}
c_{j+2a_{y},\downarrow}-c^{\dag}_{j,\uparrow}
c_{j-2a_{y},\downarrow} \nonumber
\\&+&3(1-i)\sum_{j}c^{\dag}_{j,\uparrow} c_{j-a_{x}+a_{y},\downarrow}
-c^{\dag}_{j,\uparrow} c_{j+a_{x}-a_{y},\downarrow}\nonumber
\\&+&3(1+i)\sum_{j}
c^{\dag}_{j,\uparrow}
c_{j-a_{x}-a_{y},\downarrow}-c^{\dag}_{j,\uparrow}
c_{j+a_{x}+a_{y},\downarrow}\nonumber
\\&+&4i\sum_{j}
 c^{\dag}_{j,\uparrow} c_{j+a_{y},\downarrow}
-c^{\dag}_{j,\uparrow} c_{j-a_{y},\downarrow}\nonumber
\\&+&4\sum_{j}c^{\dag}_{j,\uparrow} c_{j+a_{x},\downarrow}
-c^{\dag}_{j,\uparrow} c_{j-a_{x},\downarrow} + h.c.],
\end{eqnarray}
where  $t=\hbar^2/2m^*a_0^2$ and $t_{SO-k^3}=-\lambda/2a_{0}^{3}$,
$a_0$ is the mesh lattice spacing, and $\vec{\delta}=\pm a_0
\hat{x},\pm a_0 \hat{y}$. The first term represents  a quenched
disorder potential and disorder is introduced by randomly
selecting the on-site energy $\epsilon_j$ in the range [-W/2,W/2].
The continuum effective  mass model  for  2DES and its
tight-binding correspondent can be found
elsewhere \cite{Sheng04,Hankiewicz04}.
\begin{figure}[t]
\includegraphics[width=2.6in]{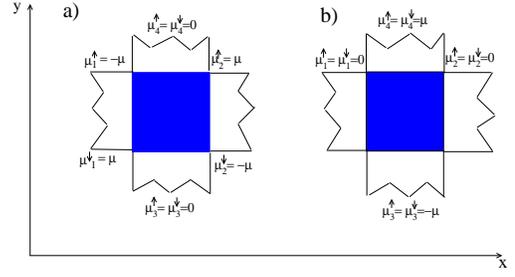}
\caption{The schematic picture of setup to measure (a)charge Hall
effect driven by spin dependent chemical potential and
(b)spin-Hall effect.} \label{fig1}
\end{figure}

\begin{figure}[t]
\includegraphics[width=3. in]{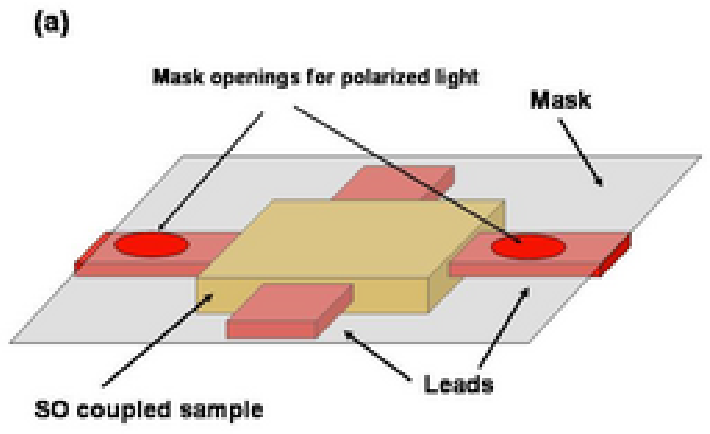}
\includegraphics[width=3. in]{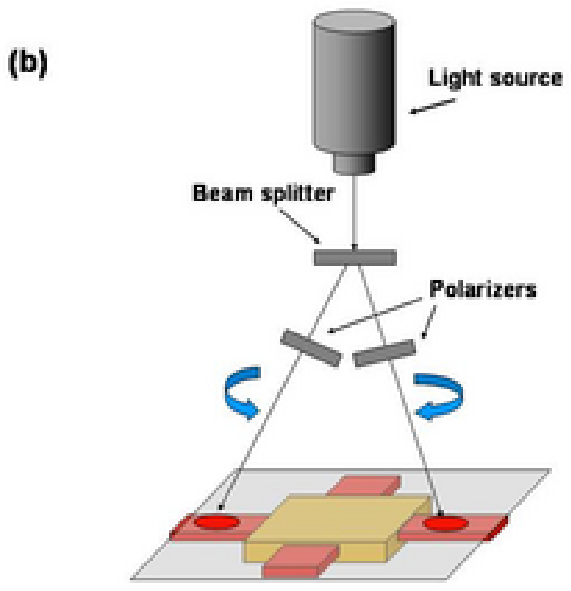}
\caption{Proposal of experimental setup to measure the reciprocal
spin-Hall effect. a) Mask (light-gray) covers a sample and leads
except for two holes in the longitudinal leads where circularly
polarized light shines. (b) Full schematics of the experimental
set-up.}
\end{figure}
 Within the leads the SO
coupling  is zero  and therefore  each lead should  be considered
as having two independent spin-channels. Moreover, in leads
without the SO coupling, the spin current is measured in a medium
where spin is conserved removing the ambiguity of spin-current
definition. For SHE, leads constitute reservoirs of electrons at
chemical potential $\mu_1,\dots,\mu_N$, where $N$ is the number of
leads that we consider  to be  four (see Fig.~\ref{fig1}b). For
RSHE, the chemical potential is spin dependent in leads 1 and 2
allowing the generation of spin-force in the x direction (see
Fig.~\ref{fig1}a). In the  low temperature limit $k_B T<<E_F$ and
for  low bias-voltage, the {\it particle} current going through a
particular channel is given within the LB formalism by
\cite{Datta}
%\begin{equation}
$I_{p,\sigma}=(e/h)\sum_{q\sigma'}T_{p,\sigma;q,\sigma'}[V_p-V_q]$,
%\end{equation}
where  $p$  labels  the   lead  and  $T_{p,\sigma;q,\sigma'}$  is
the transmission  coefficient  at  the  Fermi  energy  $E_F$
between  the $(p,\sigma)$ channel and  the $(q,\sigma')$ channel.
This transmission coefficient     is     obtained    by
$T_{p,\sigma;q,\sigma'}={\rm
Tr}[\Gamma_{p,\sigma}G^R\Gamma_{q,\sigma'}G^A]$ where
$\Gamma_{p,\sigma}$ is given by
$\Gamma_{p,\sigma}(i,j)=i[\Sigma^R_{p,\sigma}(i,j)-\Sigma^A_{p,\sigma}(i,j)]$.
The retarded and advance  Green's function of the sample $G^{R/A}$
with   the  leads  taken   into  account   through  the self
energy $\Sigma^{R/A}_{p,\sigma}(i,j)$ has a form
$G^{R/A}(i,j)=[E\delta_{i,j}-H_{i,j}-\sum_{p,\sigma}\Sigma^{R/A}_{p,\sigma}(i,j)]^{-1}$.
Here the position  representation of the matrices
$\Gamma_{p,\sigma}$, $G^R$, $H_{i,j}$, and $\Sigma^R$ are in the
subspace of the sample. Within the  above formalism the  spin
current through each channel is given  by $I^s_{p,\sigma}=
(e/4\pi)\sum_{q\sigma'}
T_{p,\sigma;q,\sigma'}[V_{p,\sigma}-V_{q,\sigma'}]$. The
spin force driven charge-Hall conductance, $G^{yx}_{CS}$, is defined as the ratio of
charge current in the y- direction induced by the spin-dependent
chemical potential along the x- axis to this spin-dependent chemical
potential difference (see Fig 1a).
\begin{equation}\label{gcharge}
G^{yx}_{CS}=\frac{(I^s_{3\uparrow}+I^s_{3\downarrow})}{V^{\uparrow}_2-V^{\uparrow}_1},
\end{equation}
where $V^{\uparrow}_i =\mu^{\uparrow}_i/e$. The spin-Hall
conductance ,$G^{xy}_{SC}$, is defined as the ratio of
spin-current in the x-direction induced by charge voltage difference
in y direction to this voltage difference
\begin{equation}\label{gspin}
G^{xy}_{SC}=\frac{(I^s_{1\uparrow}-I^s_{1\downarrow})}{V_4-V_3},
\end{equation}
where $V_i =\mu_i/e$, and the labels are indicated in
Fig.~\ref{fig1}b.  $G^{yx}_{SC}$ and $G^{xy}_{CS}$ are defined by
analogy. We set the absolute value of voltage for spin and spin force driven charge-
Hall effects as $V=\mu/e=2.5$mV.

\begin{figure}[tbh]
\includegraphics[width=3.4in]{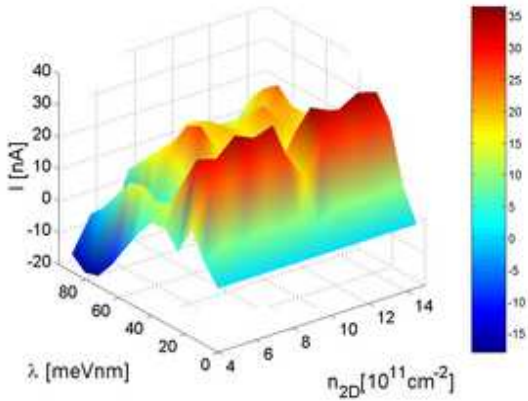}
\caption{The spin-force driven  charge current as a function of electron
concentration and Rashba coupling ,$\lambda$, for mesoscopic
square sample 100nm by 100nm and $\mu=$250,000 cm$^2$/Vs.}
\end{figure}

{\it Results  and discussion.}  In order  to address the  key
issue of experimental observation of spin driven charge-Hall
effect  as well as to establish the Onsager relation between
spin-Hall effect and it reciprocal correspondent we choose
realistic parameters for our calculations which model currently
attainable systems. We consider an effective  mass of $m^*
=0.05m_e$
 for electron systems and $m^*=0.5m_e$ for hole ones.
The disorder strength $W=0.09$  meV corresponds to  the mobility
of 250,000 cm$^2$/Vs, which is typical  for a semiconductor like
(In,Ga)As. We take the Rashba parameter $\lambda$ in the range
from  0  to 100  meV~nm, easily obtained in experiments
\cite{Koga02,Johnston02},   and we choose   the electron
concentration $n_{2D}$ in a range between $3\cdot
10^{11}$cm$^{-2}$ and $1.3\cdot 10^{12}$ cm$^{-2}$. The Fermi
energy is obtained from the chosen carrier concentration assuming
an infinite 2D gas.
\begin{figure}[tbh]
\includegraphics[width=3.4in]{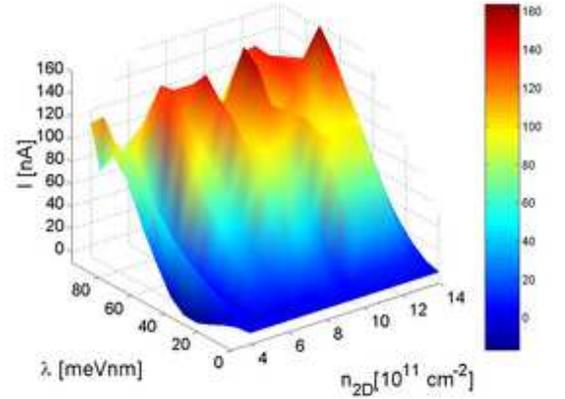}
\caption{The spin-force driven charge current as a function of hole concentration
and Rashba coupling ,$\lambda$, for mesoscopic square sample 100nm
by 100nm and $\mu=$250,000 cm$^2$/Vs.}
\end{figure}
In the detection of spin force driven charge-Hall effect the
first task is to generate a spin force which  can be
realized by spin dependent chemical potential in the leads. The
ferromagnetic leads are not the good candidates because although
 magnetization exists in ferromagnetic leads, the
chemical potential is the same for both spin directions. Here, we
propose the optical method of spin-dependent chemical potential
generation by shining the beam of circularly polarized light on
the leads (see Fig. 2). The right-circularly polarized beam shines on the right
lead (lead 2 in Fig.1a) and the left circularly polarized beam on
the left one (lead 1 in Fig.1a). The sample as well as the
transverse leads  should be covered by mask,
preventing the light absorption anywhere except the small part of longitudinal leads as shown in Fig.2a.
Choosing  semiconductor leads
 for this setup e.g. GaAs, will cause the opposite spin polarizations in left and right leads
through optical selection rules. Using the beam splitter should produce the same light intensity
 in each leads providing simultaneously the spin-dependent chemical potential between leads 1 and 2 and
the total charge current across a sample equal zero. Having
produced the spin-dependent chemical potential in the leads, we
perform calculations using non-equilibrium Green function method
presented in previous section. Fig.3 and Fig.4 present the charge
current $I_3=I^s_{3\uparrow}+I^s_{3\downarrow}$ (see sample
configuration Fig.~\ref{fig1}a) as a function of SO coupling
$\lambda$ and electron or hole concentrations, respectively. The
charge current for electron systems show oscillations with respect
to the electron density and SO coupling. The period of current
oscillations depends on system size, however its maximal value
seems to be around 40nA for systems sizes achievable in our
calculations. For hole mesoscopic systems (see Fig.4), the charge
current  also oscillates with $\lambda$ and for system sizes on
the order of 100 nanometeres is on the order of hundred
nano-ampers in a wide range of densities starting from 6$\cdot
10^{11}$cm$^{-2}$. Hence the charge current is much larger in hole
systems and shall be detectable in experiments.
\begin{figure}[h]
\includegraphics[width=3.45in]{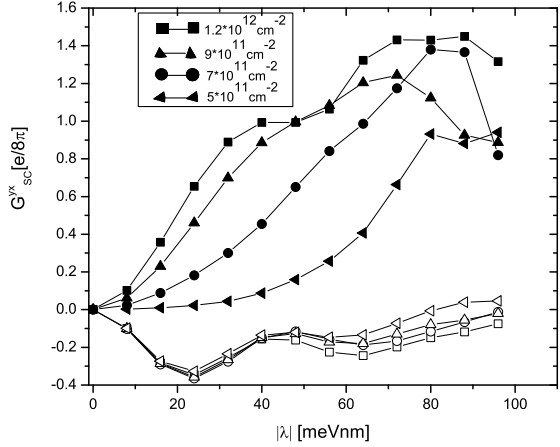}
\caption{The spin-Hall conductance as a function of SO coupling
for 100nm by 100nm square system for different concentration of
holes (close symbols) and electrons (open symbols).}
\end{figure}
 Fig.5 presents the spin-Hall conductance $G^{yx}_{SC}$
 as a function of $\lambda$ for electron and hole systems.
 One can see that spin-Hall conductance (Fig.5) and spin-force driven charge conductances (see Figs.3 and 4)
 behave similarly.
Spin-Hall conductances for electron  and hole systems oscillate
with SO coupling. Moreover, $G^{yx}_{SC}$ is several times larger
for hole systems in comparison with electron ones, which is
associated with much larger effective mass in a case of hole
systems.
 Let us emphasize that for electron and hole systems with the same effective  masses
the spin-Hall effect can be larger for electron systems. Moreover,
 in mesoscopic systems where Fermi energy as well as  multichannel effects are important
the straightforward renormalization of effective mass of electron and hole systems to  compare the spin-Hall
 conductance suggested by \cite{Wu05} does not have to be correct.
Our calculations in  mesoscopic systems are in agreement with the
linear response Kubo calculation for 2DEG and 2DHG which show that
spin-conductance is much more larger for hole systems
 \cite{Schliemann05,sinova04}. The sign of spin-Hall and charge Hall conductances
in mesoscopic systems depends on the Fermi energy.

\begin{figure}[h]
\includegraphics[width=3.4in]{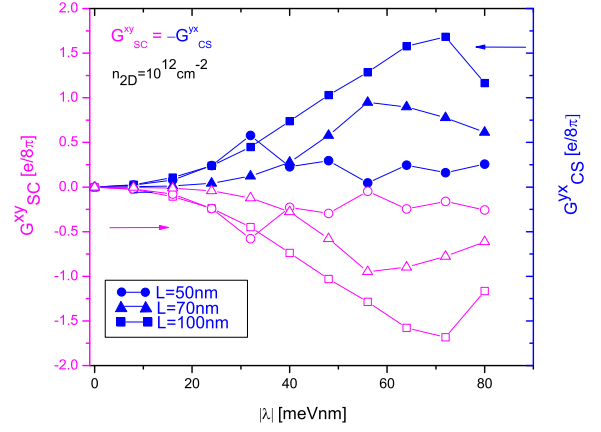}
\caption{Onsager relation between the spin-Hall conductance
$G^{xy}_{SC}$ (open symbols) and spin-force driven charge-Hall
conductance $G^{yx}_{CS}$ (close symbols) for different L by L
hole systems.}
\end{figure}

The Onsager relations express the symmetry between the transport
coefficients describing reciprocal processes in systems with a
linear dependence of response to the driving forces. Within the
models studied, considering disorder averaged conductances, the
relations can be derived by utilizing the time reversal symmetry
and the inversion symmetry on the x-y plane which relate
$T_{\sigma n;\sigma m}= T_{\bar{\sigma} m;\bar{\alpha} n}$ and
$T_{\sigma n; \alpha m}= T_{\sigma m; \alpha n}$ respectively,
where $\sigma, \alpha$ and $n,m$ are the spin and lead labels and
a bar represents the opposite direction. Within the SHE
$V_1^\sigma=V_2^\sigma=0$ and $V_3^\sigma=-V_4^\sigma=V_0/2$
following the labels of Fig. 1. Within the RSHE
 $V_3^\sigma=V_4^\sigma=0$ and $V_1^\sigma=V_2^{\bar{\sigma}}=s(\sigma)V_0/2$ where $s(\uparrow)=+1$
and $s(\downarrow)=-1$. Within the LB formalism and these boundary conditions we obtain
for the spin-current associated with the SHE
$I_1^{spin}\equiv I_{1}^{ \uparrow}-I_1^\downarrow=\frac{V_0}{2}\sum_{\sigma,\alpha}
s(\sigma)(T_{\sigma 1; \alpha 3}-T_{\sigma 1; \alpha 4})$ and for the charge current associated
with the RSHE $I_3^{c}\equiv I_{3}^{ \uparrow}+I_3^\downarrow=\frac{V_0}{2}\sum_{\sigma,\alpha}
s(\alpha)(T_{\sigma 3; \alpha 1}-T_{\sigma 3; \alpha 2})$. The above symmetries imply
that $T_{\sigma 1; \alpha 4}=T_{\sigma 2; \alpha 3}=T_{\bar{\sigma} 3;\bar{\alpha} 2}$. This
then yields $I_1^{spin}=-I_3^c$ which implies the Onsager relation
\begin{equation}\label{Onsager}
G^{xy}_{SC}=-G^{yx}_{CS}
\end{equation}
This is verified numerically in
Fig.6 which presents the disorder averaged spin-Hall conductance and charge Hall conductance for
hole systems of different sizes. For a specific disorder realization this relation does not hold
and is only approximate depending on the strength of the fluctuations induced by disorder.
This relation between $G_{SC}$ and $G_{CS}$ is consistent with predictions of semiclassical
wave-packet theory, where standard definition of spin-current was
modified by a spin-torque term \cite{Niu04,Niu05}.
However, as seen from the above derivation and noted in the introduction, our
finding of an Onsager relations in mesoscopic coherent systems do not involve such
spin-torque term since all spin-currents are defined in the non-spin-orbit coupled leads.

{\it Summary.} We have analyzed the spin Hall effect as well as
its reciprocal effect in mesoscopic hole and electron systems. We
have shown that the spin-Hall as well as the spin-dependent
chemical potential gradient driven charge-Hall conductances are
several times larger for hole systems. Further we have proposed
the experimental setup to detect the transverse charge current
driven by the spin-dependent chemical potential gradient through
transport measurements. We have shown that this charge current is
of the order of hundred nano-ampers in hole systems and should be
detectable. Also, we have established a direct relation between
the disorder average spin-Hall conductances and their reciprocal in
mesoscopic systems.

{\it Acknowledgments} We thank A. H. MacDonald for stimulating
discussions. The work was partly supported by the Research Grant
Council of Hong Kong (SQS) and DOE grant DE-FG03-02ER45958.

\bibliography{SpinHall0519}

\end{document}